\documentclass{pasj01}
\draft 
\Received{$\langle$reception date$\rangle$}
\Accepted{$\langle$acception date$\rangle$}
\Published{$\langle$publication date$\rangle$}

\begin{document}

\title{A Possible Origin of KHz QPOs in Low-Mass X-ray Binaries}
\author{Shoji \textsc{Kato}\altaffilmark{1}}
\thanks{2-2-2 Shikanodai-Nishi, Ikoma-shi, Nara 630-0114, Japan}
\altaffiltext{1}{2-2-2 Shikanodai-Nishi, Ikoma-shi, Nara 630-0114, Japan}
\email{kato.shoji@gmail.com}
\author{Mami \textsc{Machida}\altaffilmark{2}}
\thanks{Department of Physics, Faculty of Science, Kyushu University, 
744 Motooka, Nishi-ku, 819-0395, Japan}
\altaffiltext{2}{Department of Physics, Faculty of Science, Kyushu University, 
744 Motooka, Nishi-ku, 819-0395, Japan}
\email{mami@phys.kyushu-u.ac.jp}

 
\maketitle

\begin{abstract}
A possible origin of kHz QPOs in low-mass X-ray binaries is proposed.
Recent numerical MHD simulations of accretion disks with turbulent 
magnetic fields of MRI definitely show the presence of two-armed spiral 
structure in quasi-steady state of accretion disks.
In such deformed disks, two-armed ($m=2$) c-mode ($n=1$)
oscillations are excited by wave-wave resonant instability.
Among these excited oscillations, the fundamental in the radial 
direction ($n_r=0$) will be the higher kHz QPO of a twin QPOs, 
and the first overtone ($n_r=1$) in the radial direction will be the 
lower kHz QPO of the twin.   
A possible cause of the twin high-frequency QPOs (HFQPOs) in BH X-ray 
binaries is also discussed.
\end{abstract}

\section{Introduction}
 
The origin of quasi-periodic oscillations (QPOs) in low-mass
X-ray binaries is one of controversial problems.
One of their possible origins will be disk oscillations.
There is, however, no definite disk oscillation model which explains why
particular modes of oscillations are excited and appear as QPOs.
One of promising mechanisms for explaining them is the warp resonant 
model (Kato 2004, 2008; Ferreira and Ogilvie 2008; Oktarian et al. 2010).
In this model, the unperturbed disk is assumed to have been deformed from an axisymmetric state by a warp. 
Through the warp, two particular  
oscillation modes are resonantly 
coupled and excited (a kind of three mode coupling).
In this resonant instability, one of the excited oscillation modes is 
axisymmetric g-mode.\footnote{The another terminology of g-mode is 
r-mode.
}
This axisymmetric g-mode oscillation is self-trapped in the inner region of disks, and is regarded as one of quasi-periodic oscillations.

In the case where disks have poloidal magnetic fields this self-trapping 
of g-mode oscillations is questioned by Fu and Lai (2009).
Dewberry et al. (2018), however, show that this sensitivity of the trapping 
to poloidal magnetic fields is weakened if toroidal magnetic fields of
moderate strength are present.
Based on this consideration, Dewberry et al. (2019a,b) developed the 
resonant model to cases of magnetized disks, focusing on 
eccentric disks as well as warped disks.

The above-mentioned warp resonant instability can be interpreted as a 
particular case of a more general resonant instability (Kato 2013, 2014).
The latter instability is called here and hereafter   
''wave-wave resonant instability in deformed disks". 
The instability condition can be expressed in a quite simple form
(Kato 2013).
For this resonant instability to work, the disk deformation required  
is not restricted only to a warp (and eccentric deformation), but also other 
types of disk deformations are allowed, if there are disk oscillation modes which 
satisfy necessary instability conditions.
A typical example is the 3 : 1 tidal instability (Kato 2013).

In a binary system the disk around the primary star is tidally deformed.
In tidally deformed disks a disk instability which is called the 3 : 1 
resonance occurs (Whitehurst 1988; Hirose and Osaki 1990; Lubow 1991).
In a viewpoint of particle approximation, the instability is a  
parametric resonance (Hirose and Osaki 1990). 
In a viewpoint of fluid systems the cause of this instability is explained 
by a mode-mode coupling of perturbations (Lubow 1991).
In the viewpoint of the wave-wave resonant instability mentioned above, 
this 3 : 1 resonance is a particular example of the instability of the above-mentioned wave-wave resonant instability in a deformed disks (Kato 2013).    

The disk deformations expected in disks surrounding compact objects  
are not limited to warped nor tidal ones.
MHD simulations of accretion disks with MRI turbulence show 
that the magnetized accretion disks are not axisymmetric in quasi-steady
states, but have 
spiral structure (Machida and Matsumoto 2008). 
Recently, much higher resolved MHD simulations of accretion disks by use of 
CANS+ (Matsumoto et al. 2019) show definitely the presence 
of spiral structures (Machida et al. in preparation).
Then, a natural question is whether high-frequency disk oscillations observed in X-ray
binaries can be explained as the oscillations excited 
by the wave-wave resonant instability through this 
disk deformation. 
The purpose of this paper is to examine this possibility and to suggest 
that kHz QPOs observed in
low-mass X-ray binaries will be c-mode oscillations excited by this 
resonance process. 
Observational evidence shows that the high-frequency kHz QPOs 
observed in neutron-star X-ray binaries often appear in pair and
their frequency changes in time are correlated (e.g., Belloni et al. 2003).
This characteristics of correlated change of the pair QPOs is
in favor of the idea that the oscillations are two-armed c-mode ones, 
as will be shown in the text. 

In this paper, we first outline the essence of the wave-wave resonant 
instability in deformed disks formulated by Kato (2013, 2016),
and summarize the conditions of resonance and resonant instability.
Then, we apply this instability to the case where the unperturbed disks are 
deformed from an axially-symmetric ones by a two-armed slowly 
rotating pattern.
The results show
that two-armed c-mode oscillations are excited, if they are reflected back 
outwards around the inner edge of the disks.  
Furthermore, we demonstrate that the magnitude of excited frequencies and 
their time correlated changes seem to qualitatively agree with those of 
twin kHz QPOs observed in low mass X-ray binaries. 

Finally, a possible origin of high frequency QPOs (HFQPOs) observed in black-hole 
binaries is briefly
discussed from a viewpoint of the wave-wave resonant instability in 
two-armed deformed disks.

\section{Outline of Wave-Wave Resonance Instability}

Before presenting an outline of the wave-wave resonant instability, we shall
briefly classify the oscillation modes in geometrically 
thin relativistic disks.
Then, we describe conditions of the resonant interaction and the resonant instability.

\subsection{Wave Modes in Disks}

Let us consider geometrically thin axisymmetric relativistic disks 
surrounding compact objects.
To describe perturbations on the disks, we introduce cylindrical coordinates
($r$, $\varphi$, $z$) whose origin is at the center of the central object 
and the $z$-axis is perpendicular to the disk plane. 

If the disk is approximated to be isothermal in the vertical direction
($z$-direction) and the small amplitude perturbations superposed on  
axisymmetric disks are local in the radial direction 
($r$-direction) in the sense that their radial wavelength
is shorter than the characteristic radial scale of disks.
Then, the dispersion relation describing the perturbations
can be written as (e.g., Okazaki et al. 1987)
\begin{equation}
       [(\omega -m\Omega)^2-\kappa^2][(\omega-m\Omega)^2-n\Omega_\bot^2]
            =k^2c_{\rm s}^2(\omega-m\Omega)^2,
\label{disp}
\end{equation}
where perturbations have been taken to be proportional to
${\rm exp}[i(\omega t-m\varphi-kr)]$,
$\omega$ is the frequency of perturbations, $m$  
the azimuthal wavenumber $(\vert m\vert =0,1,2,3,...)$,
$k$ the wavenumber in the radial direction. 
$\Omega(r)$ is the angular
velocity of disk rotation, taken to be Keplerian one $\Omega_{\rm K}(r)$.
In equation (\ref{disp}), $\kappa$ and $\Omega_\bot$ are horizontal and vertical epicyclic frequencies, respectively, and $c_{\rm s}$ is 
the acoustic speed in disks.  
In the Newtonian disks, $\kappa$ and $\Omega_\bot$ are equal to
$\Omega_{\rm K}$, but in relativistic ones we have $\kappa(r)<\Omega_\bot(r)
<\Omega_{\rm K}(r)$ (e.g., Kato 2001).
The symbol $n(=0,1,2,3,...)$ in equations (\ref{disp}) represents the node 
number of perturbations in the vertical direction.
If $n=0$, the oscillations have no mode in the vertical direction in the sense
that there is approximately no vertical motion, i.e., the oscillations are
nearly horizontal.
In the case of $n=1$, there is vertical motion crossing the equatorial plane.

It is noted that we consider nonself-gravitationg relativistic disks, but
the effects of general relativity are taken into account only by adopting 
general relativistic expressions for $\Omega_{\rm K}$, $\kappa$,
and $\Omega_\bot$.
In others, Newtonian equations are adopted.

The oscillation modes are classified by $m$ and $n$.
The dispersion relation (\ref{disp}) gives two solutions with 
respect to $(\omega-m\Omega)^2$.
In the case of $n=0$, however, one of them is trivial, i.e.,   
$(\omega-m\Omega)^2=0$, and the other is 
\begin{equation}
        (\omega-m\Omega)^2=\kappa^2+k^2c_{\rm s}^2.
\label{disp2}
\end{equation}
This is the dispersion relation for inertial acoustic oscillations.
In the case of $n\geq 1$, since $\Omega_\bot$ is always larger than $\kappa$,
the two solutions with respect to $(\omega-m\Omega)^2$ of dispersion relation 
(\ref{disp}) are characterized by 
\begin{equation}
      (\omega-m\Omega)^2<\kappa^2
\label{disp3}
\end{equation}
and 
\begin{equation}
      (\omega-m\Omega)^2>n\Omega_\bot^2.
\label{disp4}
\end{equation}
The former mode is g mode (it is also called r-mode), while the
latter is called c-mode (corrugation mode) when $n=1$, and 
vertical p-mode when $n\geq 2$.\footnote{In many previous papers written by Kato the latter mode with $n=1$ are all called vertical p-mode
except when $m=1$;  i.e., 
only the mode with $n=1$ and $m=1$ was called c-mode. 
Here, however, all the latter mode with $n=1$ is called c-mode irrespective
of the value of $m$.
The terminology of vertical p-mode is used for modes of $n\geq 2$. 
This classification has been used in Kato (2016).}

\subsection{Conditions of Resonant Interaction}

Let us assume that the unperturbed disks are deformed from an axisymmetric 
equilibrium state by a wavy perturbation.
The perturbation is assumed to be time-periodic with frequency 
$\omega_{\rm D}$ and azimuthal wavenumber $m_{\rm D}$.
On such deformed disks, two small-amplitude normal mode oscillations are 
superposed.
The set of frequency and azimuthal wavenumber, $(\omega, m)$, of these two normal mode oscillations are $(\omega_1, m_1)$ and $(\omega_2, m_2)$.
If there are the following resonant relations:
\begin{equation}
      \omega_1+\omega_2+\omega_{\rm D}=0,\quad
      m_1+m_2+m_{\rm D}=0,
\label{resonance1}
\end{equation}
the two oscillations with $(\omega_1, m_1)$ (called hereafter mode 1)
and $(\omega_2, m_2)$ (called hereafter mode 2) resonantly interact each 
other through the disk deformation with $(\omega_{\rm D}, m_{\rm D})$
(called mode D).
This is a kind of three mode resonant coupling.

One more relation is necessary for the resonant coupling.
In idealized disks with vertically isothermal stratification, 
the variable $h_1(\equiv p_1/\rho_0)$ ($p_1$ is pressure perturbation and 
$\rho_0$ is unperturbed density in disks)
associated with disk oscillations with
vertical node number $n$ 
is proportional to ${\cal H}_n(z/H)$, i.e., $h_1\propto {\cal H}_n(z/H)$ (e.g., Kato 2001, 2016), 
where ${\cal H}_n$ is the
Hermite polynomial of index $n$ with argument $z/H$, $H$ being the vertical scale length of disks.
If an oscillation mode with $n_1$ nonlinearly couples with a disk deformation 
with $n_{\rm D}=0$,
the resulting mode is an oscillation with $n_1$.
If the disk deformation has $n_{\rm D}=1$, however, the oscillation resulting 
from the nonlinear coupling is 
${\cal H}_{n\pm 1}(z/H)$.
Hence, for mode 1 and mode 2 to couple each other through a disk
deformation of $n_{\rm D}=1$, the relation
$n_2=n_1+1$ or $n_2=n_1-1$ is necessary.
In summary, the additional resonant condition is 
\begin{eqnarray}
                &&n_2= n_1; \quad \hspace{108pt}{\rm when}\quad n_{\rm D}=0
                       \nonumber \\
                &&n_2=n_1+1 \quad {\rm or} \quad n_2=n_1-1;
                        \quad {\rm when}\quad n_{\rm D}=1,
\label{resonance4}
\end{eqnarray}
where $n_1$, $n_2$, and $n_{\rm D}$ are the vertical node numbers of
mode 1, mode 2, and the mode of disk deformation, respectively. 

Finally, it should be noted that for three modes (mode 1, mode 2 and 
mode D) to have a nonlinear coupling, their radial propagation regions 
should be overlapped.
Otherwise, there is no nonlinear coupling.  
   
\subsection{Conditions of Resonant Instability}

All conditions mentioned in subsection 2.2 are necessary for resonant 
coupling among three modes.
We assume that the disk deformation mode (mode D) has a large amplitude compared with
other two disk oscillation modes and is maintained by some excitation 
processes which are not considered here.
Then, two modes (mode 1 and mode 2) are resonantly amplified or damped
by the resonant coupling.
The condition of resonant amplification is found to be written 
in a simple and general form.
That is, the condition of amplification is (Kato 2013, see also Kato 2016)
\begin{equation}
     \biggr(\frac{E_1}{\omega_1}\biggr)\biggr(\frac{E_2}{\omega_2}\biggr)>0,
\label{condition1}
\end{equation}
where
$E_1$ and $E_2$ are wave energies of mode 1 and mode 2, 
respectively.\footnote{An expression for wave energy $E$ for normal mode
oscillations is presented, for example, in Kato (2001, 2016).
A simple expression for $E$ is equation (3.31) in Kato (2016).
}
This instability condition is quite general and valid even in magnetized
disks (Kato 2014).

\section{Two-Armed Deformed Disks and Excitation of Two-Armed C-mode Oscillations} %

For the wave-wave resonant instability outlined in section 2 to work, 
the unperturbed disks need to be deformed. 
Here, we consider the case where the disk is deformed by two-armed
pattern.
The wave-wave resonant instability known up to the present by
other types of disk deformation 
will be briefly summarized in Appendix.

\subsection{Presence of Two-Armed Pattern in Disks}

Recently, high-resolution numerical MHD simulations of accretion disks 
have been done (Machida et al. in preparation), where 
angular momentum is transported by turbulence of 
magneto-rotational instability (MRI).
The pseudo-Newtonian potential is used.
The initial states of the systems are almost the same as those adopted by
Machida et al. (2003, 2006).\footnote{In Machida et al. (2003) the center
of the initial torus was at $50\, r_{\rm S}$, while in the new simulations
it is at $40\, r_{\rm S}$.
In Machida et al. (2006) the temperature of the initial torus was slightly 
lower than that in the new simulations.
}
That is, the initial disk is a torus with horizontal magnetic fields
of 100 plasma $\beta$ ($\beta\equiv p_{\rm gas}/p_{\rm mag}$).
The numbers of mesh points in the radial, azimuthal, and vertical directions
are $(N_r,N_\varphi, N_z)=(774, 512, 774)$ in the maximum cases.
The mass of the central source is $10M_\odot$.

\begin{figure}
 \begin{center}
        \includegraphics[width=120mm]{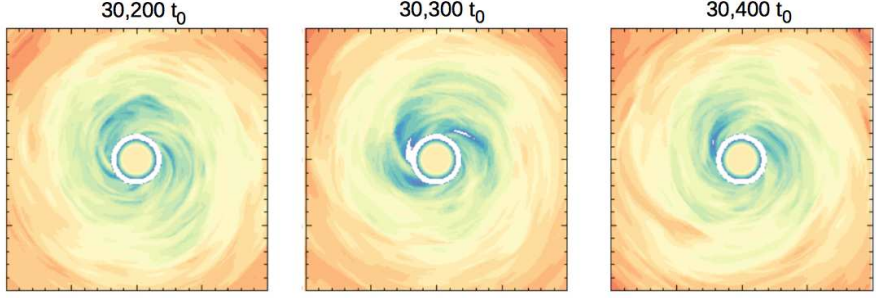}
 \end{center}
 \caption{Density distribution on the equatorial plane at epochs of 
 $t=3.02\times 10^4 t_0$, $t=3.03\times 10^4 t_0$, $t=3.04\times 10^4 t_0$,
 where $t_0$ is the unit time defined by the Schwarzschild radius $r_{\rm S}$ 
 and the speed of light $c$, i.e., $t_0=r_{\rm S}/c$.
 The radius of the central circle is $3r_{\rm S}$.
 Comparison of the three panels shows that the spiral pattern rotates
 slowly in the prograde direction.    
  (calculated from the data of Machida et al 2020)}
 \label{.....}
\end{figure}
\begin{figure}
 \begin{center}
        \includegraphics[width=120mm]{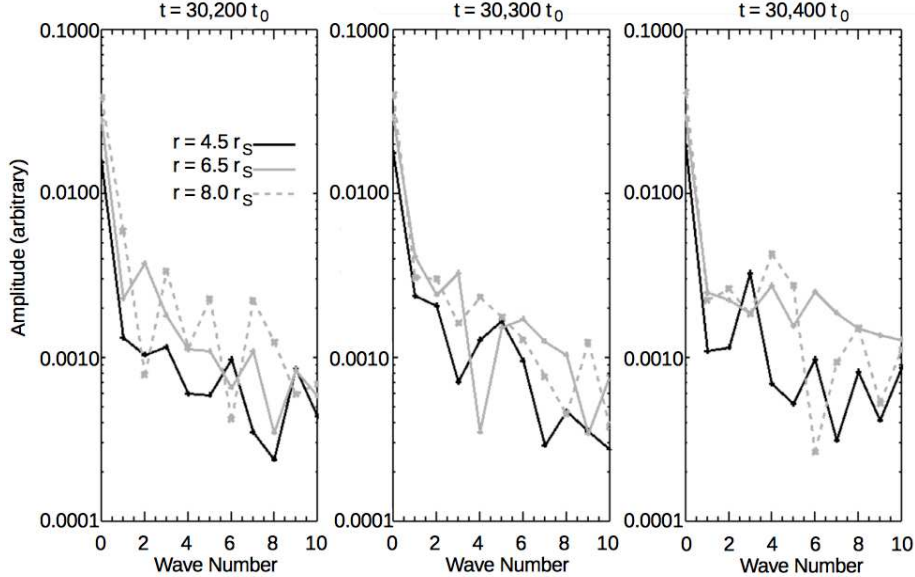}
 \end{center}
 \caption{The Fourier components of density variations in the azimuthal direction 
 at radii of $r=4.5r_{\rm S}$, $6.5r_{\rm S}$, and $8.0r_{\rm S}$.
 The density averaged in the azimuthal direction is normalized to be unity.
 The abscissa denotes azimuthal wavenumber $m$.
 (calculated from the data of Machida et al 2020)
 } 
 \label{.....}
\end{figure}

The simulation results clearly show that the disks in the quasi-steady state 
are not axisymmetric, but spirally deformed.
Figure 1 shows an example of density distributions  
on the equatorial plane with less spatial resolution of mesh points
(774, 64, 774).
Three epochs of $3.02\times 10^4 t_0$,
$3.03\times 10^4 t_0$, and $3.03\times 10^4 t_0$ are shown,
where $t_0$ is the unit time defined by the Schwarzschild radius $r_{\rm S}$ and the speed of light $c$, i.e., $t_0=r_{\rm S}/c$.
Comparison of the above three panels seems to show that the spiral pattern 
rotates slowly 
in the direction of disk rotation, but 
the angular velocity of the pattern rotation is much slower than that of
the disk rotation. 
The spiral pattern has many arms.
In order to see the strength of each arm, 
the density variation in the azimuthal direction is 
Fourier-analyzed at a few radius. 
The results are shown in figure 2 for $r=4.5r_{\rm S}$, $6.5r_{\rm S}$,
and $8.0r_{\rm S}$.
This figure shows that the amplitude of spiral density pattern over
the averaged one is roughly a few to ten percent in density.
The pattern of $m=2$ is certainly present, 
although it is not the most dominant one.
What we need here is the $m=2$ component.\footnote{
In the present excitation model of kHz QPOs, the presence of a low-frequency 
two-armed pattern is essentially necessary.
The presence of other components has no effects on our model, 
because they do not contribute to the resonance.
}

The origin of spiral disk deformation is an interesting subject, but
here we simply assume that on the unperturbed disks a plane-symmetric, 
low-frequency two-armed pattern is present and maintained.
That is, we assume the presence of a low-frequency, two-armed spiral pattern
of 
\begin{equation}
        m_{\rm D}=-2, \  \quad n_{\rm D}=0, \quad {\rm and} \quad 
        \omega_{\rm D}\sim 0,
\label{deformation}
\end{equation}
where we have adopted $m_{\rm D}=-2$ (not $m_{\rm D}=2$) for convenience, 
because there is no essential difference between waves of 
${\rm exp}[i(\omega_{\rm D}t-m_{\rm D}\varphi)]$ 
and those of ${\rm exp}[-i(\omega_{\rm D}t-m_{\rm D}\varphi)]$.

\subsection{Mode 1 and Mode 2 Satisfying Resonant and Excitation Conditions}

As two waves which resonantly interact through the above disk deformation,
we consider the following set of two oscillations modes:
\begin{eqnarray}
     &&{\rm mode}\ 1: \quad  {\rm c}-{\rm mode} \ {\rm oscillations\ with}\ 
                           (m_1=2 \ ,\ n_1=1, \ \omega_1), 
\label{resonance 2}               \\
     &&{\rm mode}\ 2: \quad  {\rm c}-{\rm mode} \ {\rm oscillations\ with}\
                           (m_2=0 \ ,\ n_2=1, \  \omega_2\sim -\omega_1).
\label{resonance 3}
\end{eqnarray}

The issue to be examined next is whether the instability condition 
(\ref{condition1}) is satisfied by the above oscillation modes.
The propagation region of mode 1 specified by equation (\ref{resonance 2})
is given by $(\omega_1-2\Omega)^2>\Omega^2_\bot$ [see inequality (\ref{disp4})], which is
divided into two regions of $\omega_1>2\Omega
+\Omega_\bot$ and $\omega_1< 2\Omega-\Omega_\bot$.
We are interested in the oscillations in the latter propagation region,
which is schematically shown in figure 3.
The oscillation in this latter propagation region has a negative wave energy, 
since the propagation region
is inside the corotation radius of $\omega=2\Omega$.\footnote{
The sign of wave energy $E$ of an oscillation with $\omega$ and $m$, 
sign($E$), is the same as
the sign of $\omega(\omega-m\Omega)$ in the propagation region of the 
oscillation (e.g. see Kato 2016), i.e., sign$(E)<0$ if $\omega/m<\Omega$,
while sign$(E)>0$ if $\omega/m>\Omega$.
In other words, we have sign$(E)<0$ inside the corotation radius of
$\omega/m=0$,
while sign$(E)>0$ outside the corotation radius of $\omega/m=0$.
} 
Hence, we have $E_1/\omega_1<0$.
Next, we consider the c-mode oscillation specified by equation 
(\ref{resonance 3}).
The propagation region of the oscillations is specified by
$\omega_2^2>\Omega_\bot^2$ [see inequality ({\ref{disp4})].
This propagation region consists of two domains, i.e., $\omega_2>\Omega_\bot$
and $\omega_2<-\Omega_\bot$.
We are interested in the oscillations in the region of 
$\omega_2<-\Omega_\bot$, which is also schematically shown in figure 3.
This oscillation is axisymmetric and thus the wave energy is 
positive, leading to $E_2/\omega_2<0$, since $\omega_2<0$.
Hence, the condition of resonant instability (\ref{condition1}) is satisfied.

It is noted that the condition of resonance in frequency, i.e., 
\begin{equation}
         \omega_1+\omega_2+\omega_{\rm D}=0
\label{frequency-cond}
\end{equation}
is satisfied by adopting $\omega_2$ so that the above 
resonant condition among frequencies holds for given $\omega_1$ and 
$\omega_{\rm D}$.
This is always possible, because mode 2 has practically no
definite outer boundary (see figure 3).  
In other words, mode 2 is not a trapped oscillation.

In the present case the signs of $\omega_1$ and $\omega_2$ are opposite,
because $\omega_{\rm D}$ is smaller than $\omega_1$.
Hence, the condition of resonant instability (\ref{condition1}) can be 
written as $E_1E_2<0$.
This has a simple physical interpretation.
That is, the resonant interaction of two oscillations with opposite
signs of wave energy leads to instability by a positive energy flowing to
the oscillation with positive energy from the one with negative energy.

\begin{figure}
 \begin{center}
  \includegraphics[width=80mm]{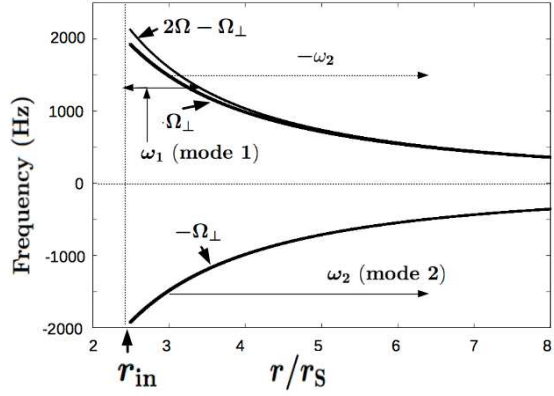}
 \end{center}
 \caption{Schematic figure showing one of propagation regions of two-armed 
 c-mode oscillations (called mode 1) and one of propagation regions of
 axisymmetric p-mode oscillations (called mode 2).
 It is noticed that the propagation region of mode 1 is limited inside by
 the inner edge of the disks, $r_{\rm in}$, and outside by the radius where 
 $\omega_1=2\Omega-\Omega_\bot$.
 That is, the oscillation is trapped and has a discrete frequency.
 Different from this, the propagation region of mode 2 is limited inside by  
 the radius where $-\omega_2= \Omega_\bot$, but there is no 
 explicit outer boundary.
 That is, the frequency of $\omega_2$ is determined passively by the 
 resonance condition. 
 The curves of $\Omega_\bot$ and $2\Omega-\Omega_\bot$ in this diagram 
 are drawn in the case of spin parameter $a_*$ is 0.3 in order to 
 demonstrate the difference of $\Omega_{\rm K}$ and $\Omega_\bot$. 
 } 
 \label{.....}
\end{figure}
  
It should be noticed that since $\omega_1$ and $-\omega_2$ are close
($\omega_{\rm D}$ is small), 
the outer edge (radius) of the propagation
region of mode 1 and the inner edge of the propagation region of mode 2
are close (see figure 3), even when they are not overlapped.
In the evanescent regions of oscillations, the amplitude of the oscillations
spatially damps but it is still large near the boundary of the propagation region. 
Consequently, the necessary condition that the propagation regions of 
mode 1 and mode 2, and the disk deformation region are 
overlapped is practically satisfied.

\section{Frequencies of Excited Oscillation Modes and kHz QPOs}     %

Here, we examine whether the set of mode 1 and mode 2 
considered above can describe the frequency of a kHz QPO.
First, we should consider the frequency $\omega_1$ of mode 1.
To obtain $\omega_1$ a boundary condition is necessary at an inner part 
of the disks.
This is, however, a difficult subject, because we do not know 
what boundary condition can be imposed at the inner edge $r_{\rm in}$ 
of disks.
(If we state more critically, it is uncertain whether we can 
impose a boundary condition near $r_{\rm in}$.)
Near the inner edge of the disks, $r_{\rm in}$, the gas density decreases
sharply towards the central source.
Hence, a simple consideration suggests that Lagrangian pressure variation
vanishes there, i.e., $\delta p=0$ at $r_{\rm in}$.
However, we should consider that the accretion disks have inward flows and
they become supersonic beyond the sonic point.
In other words, the reflected waves are swallowed to the central object 
unless their outward wave velocity is sonic or more.  
A measure whether the reflected wave can go actually outwards against the 
inward accretion flow will be to compare the group velocity of
the c-mode waves at $r_{\rm in}$, $c_{\rm c}$, with acoustic speed 
$c_{\rm s}$.
A rough estimate of $c_{\rm c}$ under the use of the local dispersion 
relation (\ref{disp}) shows that $c_{\rm c}$ is smaller than $c_{\rm s}$
near $r_{\rm in}$.
This suggests that there is no reflection of the c-mode oscillations at
$r_{\rm in}$.

However, we should consider following situations.
The disk thickness $H(r)$ near $r_{\rm in}$ decreases sharply towards 
$r_{\rm in}$.
That is, $dH(r)/dr<0$ with a large negative value.
A direct effect of this decrease of disk thickness on wave motions will be
their reflection outwards before approaching to $r_{\rm in}$.
This reflection may occur favorably for c-mode oscillations, 
because c-mode oscillations extend in the vertical direction,
i.e., $p_1/\rho_0\propto {\cal H}_1(z/H)\propto (z/H)$ in the case of local 
approximations with $dH/dr=0$, where ${\cal H}_n$ is the Hermite polynomial 
with index $n$.
(It is noted that  
$p_1/\rho_0\propto {\cal H}_0(z/H)\propto (z/H)^0$ in p-mode oscillations.) 
It is further noticed that the c-mode oscillations are nearly incompressible
motions, i.e., $\delta\rho\sim 0$ (see e.g., Kato et al. 2008 or Kato 2016), 
and thus will be reflected on
a discontinuous surface. 
Second, the effects of sharp decrease of disk thickness 
are not considered in deriving 
the dispersion relation (\ref{disp}). 
The sharp decrease of disk thickness and strong (MRI) turbulence will
introduce modification of oscillation modes and their couplings, which will brings about reflection of oscillations.\footnote{
In the case of p-mode oscillations, wave reflection (and instability)
near the sonic point is known (e.g., Matsumoto et al. 1988; Kato et al.
1988; Honma et al. 1992).
This is related to adoption of a diffusion-type viscous force
(information can be transported through the sonic point outwards) or
topology around the sonic point (e.g., nodal-type critical point),
and might not be directly related in the present subject.
}
Taking the above situations into account, we impose in this paper a 
boundary condition at
$r_{\rm in}$, which is $p_1/\rho_0=0$ for simplicity.\footnote{
For comparison, a case where $\partial (p_1/\rho_0)/\partial r=0$ is adopted
as the inner boundary condition has been examined (Kato 2011).
The value of $\omega_1$ is quantitatively different from that in the case of
$p_1/\rho_0=0$ (see figure 5 in Kato 2011), but there is no qualitative
difference in the trend on the frequency-frequency curve given in figure 5 below.
}

By imposing this condition as the inner boundary condition 
and by imposing the radius where $\omega_1=2\Omega-\Omega_\bot$ as a turning
point of the wave equation, Kato (2012b) solved the local wave equation\footnote{
This wave equation is the one derived under approximation of a constant 
disk thickness, and corresponds to the dispersion relation (\ref{disp}).
}
describing oscillations by the WKBJ method to estimate the frequency 
of mode 1.
The frequency $\omega_2$ is determined by the resonance condition of
$\omega_1+\omega_2+\omega_{\rm D}=0$.
It is important to note here that mode 2 will not be
observed clearly, since the oscillation is an outgoing running wave 
(see figure 3)
(not a trapped one) and will be phase-mixed.
That is, the oscillation of $\omega_1$ alone is observed as a kHz QPO.
This situation is the same as that in the case of superhumps (see Kato 2013).

The next issue is why kHz QPOs are often observed in pairs.
It should be emphasized that the trapped c-mode oscillation with $m_1=2$ 
and $n_1=1$ (mode 1) has overtones resulting from  
node number(s) in the radial direction. 
The fundamental oscillation in the radial direction has no node in the radial 
direction, i.e., $n_r=0$.
This oscillation will correspond to the higher frequency QPO of a twin, 
and the lower frequency
QPO will be the first overtone oscillation with $n_r=1$.
This idea that the two-armed c-mode oscillations with $n_r=0$ and $n_r=1$ 
are the pair of kHz QPOs is already suggested by Kato
(2012b), but at that time we have no definite idea concerning excitation of these modes.
Figure 4 demonstrates the trapping of c-mode oscillations with $n_r=0$ and $n_r=1$
on the propagation diagram in the case of a non-rotating central object.

\begin{figure}
 \begin{center}
           \includegraphics[width=80mm]{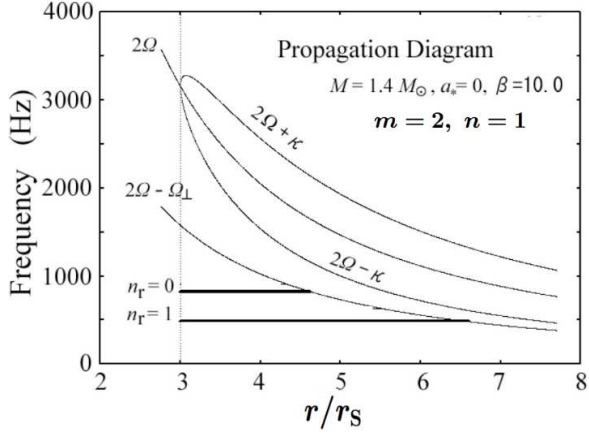}
 \end{center}
 \caption{Propagation diagram for two-armed c-mode oscillations ($n=1$)
 with $n_r=0$ and $n_r=1$.
 The horizontal lines show the radial range where the oscillations are 
 trapped.
 The disk temperature $c_{\rm s}$ is taken so that 
 $\beta[\equiv c_{\rm s}^2/ 
 (c_{\rm s}^2)_0]=10.0$ [see equation (\ref{cs}) for $(c_{\rm s}^2)_0$].
 $M=1.4M_\odot$ and $a_*=0$ are adopted.
 (reproduced from Kato 2016) 
 } 
 \label{.....}
\end{figure}

Here, the main results by Kato (2012b) are presented.  
In his calculations, the disk temperature was taken as a parameter.
As the reference temperature he adopted the temperature in the standard 
disks where 
gas pressure dominates over radiation pressure and opacity comes from the free-free processes.
Furthermore, the conventional viscosity parameter, $\alpha$, and the mass accretion rate normalized by the Eddington critical accretion rate are
taken, respectively, to be 0.1 and 0.3.
That is, as the reference isothermal acoustic speed, $[c_{\rm s}(r)]_0$, 
he adopted
\begin{equation}
      [c_{\rm s}^2(r)]_0=1.79\times 10^{16}\biggr(\frac{M}{M_\odot}\biggr)^{-1/5}\biggr(\frac{r}{r_{\rm S}}\biggr)^{-9/10} 
        \ {\rm cm}^2\, {\rm s}^{-2},
\label{cs}
\end{equation}
where $M$ is the mass of the central source and $r_{\rm S}$ is the
Schwarzschild radius, defined by $r_{\rm S}=2GM/c^2$.

Kato adopted $\beta_{\rm s}[\equiv c_{\rm s}^2/(c_{\rm s}^2)_0]$ as a dimensionless parameter for describing the disk.
The value of $\beta_{\rm s}$ will change in time by changes of accretion 
rate $\dot M$ and viscosity parameter $\alpha$.
Since the trapped region of oscillations is narrow in the radial direction,
$\beta_{\rm s}$ is taken to be a constant parameter independent of $r$.
For some values of parameter $\beta_{\rm s}$, the frequencies of 
trapped oscillations for the fundamental mode in the radial direction ($n_r=0$) and those for the first overtone ($n_r=1$) have been calculated.
An example of his calculated results is shown in figure 5, which is a 
reproduction of figure 5 by Kato (2012b).
This figure shows the correlated frequency change 
(by the change of $\beta_{\rm s}$) of the fundamental
($n_r=0$) and the first overtone ($n_r=1$) on the frequency-frequency
diagram.
On this diagram frequencies of the observed higher and lower kHz QPOs 
are over-plotted. 
This figure shows that the calculated values of $\omega$ for $n_r=0$ and $n_r=1$
are in the observed frequency range of twin kHz QPOs
and also their changes by change of $c^2_{\rm s}$ qualitatively agree 
with the observed correlated change of twin QPOs.

\begin{figure}
 \begin{center}
           \includegraphics[width=80mm]{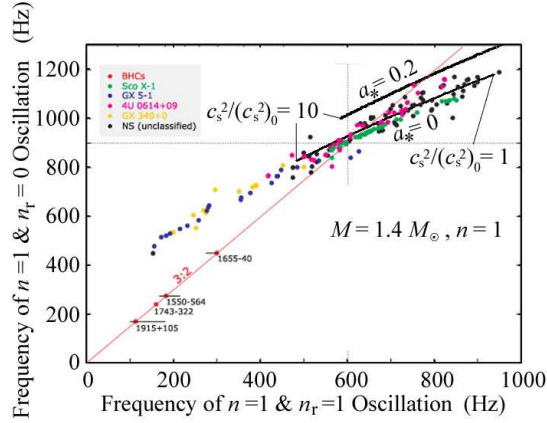}
 \end{center}
 \caption{Frequencies and their correlation between two-armed c-mode ($n=1$)
  oscillations with $n_r=0$ and $n_r=1$ for two cases of $a_*=0$ and $0.2$.
  The value of $c_{\rm s}^2/(c_{\rm s}^2)_0$ is changed from 1.0 to 10.0.
  $M=1.4M_\odot$ is adopted.
  The calculated curves are superposed on a figure by Abramowicz (2005)
  showing observational data on the frequency-frequency diagram.
  (The ordinate is the lower frequency of twin QPOs, and the abscissa is
  the higher frequency of the twin QPOs.) 
  Original plots of observational data are those by Belloni et al. (2005).  
  [Reproduction of Figure 5 in Kato (2012) ]}
 \label{.....}
\end{figure}

Finally, the effects of magnetic fields on the above arguments are
mentioned.
Even if magnetic fields are present in disks, the expression for the 
wave-wave resonant 
instability condition, $(E_1/\omega_1)(E_2/\omega_2)>0$, is kept 
unchanged (Kato 2014), although $E_1$ and $E_2$ are wave energies where
the effects of magnetic fields are taken into account.
In addition, in the cases where magnetic fields in the unperturbed state are horizontal, 
the behaviors of c-mode oscillations on the frequency-frequency diagram 
qualitatively agree with observations (Kato 2012a).
That is, Kato (2012a) examined 
the effects of magnetic fields in the horizontal direction on the 
c-mode oscillations with $m=2$.
In this study the magnetic fields in the horizontal direction 
are assumed to be stratified in the vertical direction so that
the Alfv\'{e}n speed is constant in the vertical direction
(i.e., $B_0^2\propto \rho_0$).
Even in such simplified cases the wave equation describing the 
c-mode oscillations 
is complicated to solve directly by the WBKJ method.
Hence, in that paper the oscillations are assumed to be nearly vertical  and the motions in the radial direction associated with the oscillations are treated as a perturbation over the vertical motions (this approximation is 
valid in the case of c-mode oscillations with short radial wavelength).
The results show that the calculated frequency-frequency correlation curve 
is qualitatively on the curve representing the trend of
observed correlation, although the frequency range obtained by the calculations
is changed from that in the case of no magnetic fields 
(for detail, see figures in Kato 2011).

\section{Summary and Discussion on HF QPOs in BH X-ray binaries}   %

Recent high-resolution MHD numerical simulations of accretion disks
with high plasma $\beta(\equiv p_{\rm gas}/p_{\rm mag}$),
starting from tori with toroidal magnetic fields, show that accretion disks 
surrounding compact objects with high plasma $\beta$ are not axially symmetric
in quasi-steady state,
but have spiral structures (Machida et al. in preparation).
Although the structure is complicated, 
spectrum analyses show a stiff presence of two-armed slowly 
progressive structure.
Elucidation of the origin of the structure is of interest, but here
we simply assume that the structure is maintained on the disks by some MHD precesses.
That is, we assume the presence of disk deformation of ($m_{\rm D}=2$,
$n_{\rm D}=0$) with a low frequency of $\omega_{\rm D}$.
Then, high-frequency, two-armed c-mode oscillations of $m=2$ and $n=1$
are excited by the wave-wave resonance instability
studied by Kato (2013, 2014), if the modes are reflected back outwards
around the inner edge of the disks.  
We suggests that their fundamental mode ($n_r=0$) in the radial direction
and the first overtone ($n_r=1$) are, respectively, the 
upper and the lower kHz QPOs observed in low-mass X-ray binaries.

A weak point of the present model of kHz QPOs is whether the c-mode
oscillations are actually reflected back near the inner edge of the disks.
Careful two-dimensional analyses of wave motions in accreting flows
will be necessary in future. 
If wave reflection actually occurs as is expected in this paper, comparisons 
of calculated frequencies and observed ones will be helpful for
knowing the structure around the innermost region of disks
(discoseismology). 

In this paper high-frequency QPOs (HFQPOs) of black hole 
sources were outside of our main interest, because their observational characteristics 
are different from those of kHz QPOs of neutron-star X-ray binaries.  
That is, the twin HFQPOs in black-hole sources have a frequency ratio close 
to 3 : 2 and are almost time-independent, unlike the kHz QPOs.
It is of interest to discuss why the HFQPOs in black-hole X-ray binaries
are different in their characteristics from those in kHz QPOs in neutron stars,
and whether they can be interpreted by the wave-wave resonant instability model. 
In the followings, we try to discuss the origin of the twin HFQPOs in the framework
of the wave-wave resonant model.

The neutron stars in low-mass X-ray binaries are generally
old, and thus the accretion disks surrounding the neutron stars 
will not have strong poloidal magnetic fields.
In the case of BH binaries, however,
the central part of accretion disks may be subject to poloidal magnetic 
fields of the central BH sources.
In such cases, the propagation 
regions of wave modes (including vertical p-modes), except for those of p-modes, are 
strongly affected by the poloidal magnetic fields (Fu and Lai 2009).  
Hence, even if the BH disks have the same kind of disk deformation as that in
NS disks (i.e., $m_{\rm D}=2$ and $n_{\rm D}=0$), it will be natural that the frequency 
ratio of the c-mode oscillation modes do not have the same characteristics as
those in the case of NS disks.

Next, we speculate the origin of the twin HFQPOs.
Observations show that the twin HFQPOs in BH X-ray binaries are observed
in the transition state from the low/hard state to the high/soft one
(Remillard 2005).
The transition state will be on the way to transit from
ADAFs with strong turbulent magnetic fields to optically thick standard disks
with low plasma $\beta$.
As mass accretion increases in ADAFs, they cannot be in equilibrium states.
They will vertically contract by thermal instability
due to radiative cooling, leading to low-$\beta$ disks with coherent strong 
magnetic fields (Machida et al. 2006).
We speculate that in such transition disks with strong poloidal and toroidal magnetic 
fields, 
the deformed pattern in disks may be plane-asymmetric, unlike the case of
high-$\beta$ disks considered in low-mass X-ray binaries.
In other words, we think that in the intermediate state of the BH X-ray binaries
the disk deformation from steady axisymmetric state will be spiral patterns
which are asymmetric 
with respect to equatorial plane (i.e., $n_{\rm D}=1$), unlike in the case of 
NS X-ray binaries where the spiral pattern will be plane-symmetric (i.e., $n_{\rm D}=0$).

In the above context, we assume that spiral disk deformations of $m_{\rm D}=2$ and
$n_{\rm D}=1$ are present in the disks of BH X-ray binaries in their intermediate state.
That is, we assume the presence of 
\begin{equation}
    m_{\rm D}=-2, \quad n_{\rm D}=1,\quad {\rm and}, 
                  \quad \omega_{\rm D}\sim 0,
\label{disk-deformation2}
\end{equation}
where we have again adopted $m_{\rm D}=-2$ instead of $m_{\rm D}=2$ for
convenience.
We then examine what types of oscillations are expected in these deformed disks 
by the wave-wave resonant processes.

The following set of oscillations satisfy the resonant conditions (\ref{resonance1}):
\begin{eqnarray}
     &&{\rm mode}\ 1: \quad  {\rm p}-{\rm mode} \ {\rm oscillations\ with}\ 
                           (m_1=2,\ n_1=0, \ \omega_1), 
\label{resonance 5}               \\
     &&{\rm mode}\ 2: \quad  {\rm c}-{\rm mode} \ {\rm oscillations\ with}\
                           (m_2=0,\ n_2=1, \ \omega_2\sim -\omega_1).
\label{resonance 6}
\end{eqnarray}
The propagation region of mode 1 is specified by 
$(\omega_1-2\Omega)^2-\kappa^2>0$.
This propagation region is divided into two domains.
We are interested in the trapped oscillations in the region of 
$\omega_1<2\Omega-\kappa$ (see figure 6).
The waves in this propagation region have negative energy, $E_1<0$, 
since the waves are inside the corotation radius of $\omega_1=2\Omega$ (see also figure 6).
Hence, we have $E_1/\omega_1<0$.
The propagation region of mode 2 is specified by $\omega_2^2-\Omega_\bot^2>0$.
We are interested in the oscillations in region of 
$\omega_2<-\Omega_\bot$ (see figure 6).  
The wave energy of this oscillation mode is positive, since the wave is axisymmetric, i.e., $E_2>0$.
Hence, we have $E_2/\omega_2<0$, since $\omega_2<0$.
This consideration shows that the above two oscillations of mode 1 and mode 2
are resonantly excited, since $(E_1/\omega_1)(E_2/\omega_2)>0$.
The propagation region of mode 2 has no definite outer boundary.
Hence, as in the case of kHz QPOs the oscillation of mode 2 will not be
clearly observed.
The oscillation of mode 1 will correspond to the lower frequency one of 
a twin HFQPOs in black-hole sources, if the oscillation mode 
is reflected back around $r_{\rm in}$.

\begin{figure}
 \begin{center}
  \includegraphics[width=80mm]{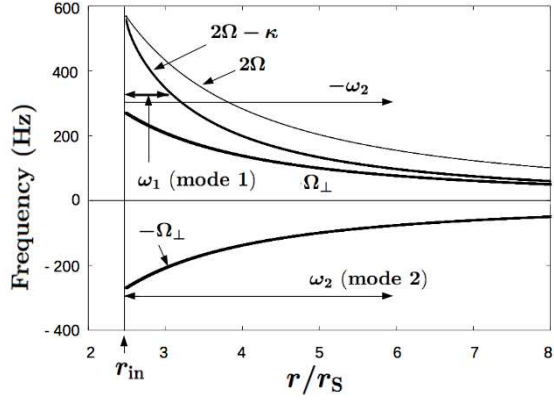}
 \end{center}
 \caption{Schematic figure showing the propagation region of a two-armed 
 p-mode oscillation ($m=2$, $n=0$, called mode 1) and the propagation region of the 
 resonantly coupled axisymmetric c-mode oscillations ($m=0$, $n=1$, called mode 2).
 The propagation region of mode 1 is limited by
 the inner edge of the disk, $r_{\rm in}$, and outside by the radius where 
 $\omega_1=2\Omega-\kappa$.
 That is, the oscillation is trapped and has a discrete frequency.
 Different from this, the propagation region of mode 2 is limited inside by  
 the radius where $-\omega_2= \Omega_\bot$ (or by $r_{\rm in}$), but there is no 
 definite outer boundary.
 That is, the frequency of $\omega_2$ is determined passively by the 
 resonance condition. 
 The curves of $\Omega_\bot$, $2\Omega-\kappa$ and $2\Omega$ in this diagram 
 are drawn in the case of spin parameter $a_*$ is 0.3. 
 The $\omega_1$ oscillation presented in this figure corresponds to the lower frequency 
 QPO of twin HFQPOs in BH X-ray binaries.
 The $\omega_1$ oscillation for the higher QPO is the p-mode oscillation of
 $m_1=3$.
 In this case the curve corresponding to $2\Omega-\kappa$ is
 $3\omega-\kappa$ and the curve corresponding to $-\Omega_\vert$ is $-\omega-\Omega_\vert$.
 The frequency of the trapped oscillation, $\omega_1$, in this case is higher than that
 shown in this figure. 
 } 
 \label{.....}
\end{figure}
  
The higher frequency oscillation of a twin 
HFQPOs in black-hole sources will be the mode 1 of the following
set of resonant oscillations under the disk deformation given by
(\ref{disk-deformation2}):
\begin{eqnarray}
     &&{\rm mode}\ 1: \quad  {\rm p}-{\rm mode} \ {\rm oscillations\ with}\ 
                           (m_1=3 \ ,\ n_1=0, \ \omega_1), 
\label{resonance 7}               \\
     &&{\rm mode}\ 2: \quad  {\rm c}-{\rm mode} \ {\rm oscillations\ with}\
                           (m_2=-1 \ ,\ n_2=1, \ \omega_2\sim -\omega_1).
\label{resonance 8}
\end{eqnarray}
Based on arguments similar to those in the case of $m_1=2$, we consider
the trapped oscillation in the region of $\omega_1<3\Omega-\kappa$ as mode 1, 
and the oscillation in the region of $\omega_2<-\Omega-\Omega_\bot$ as mode 2. 
The above-mentioned two modes satisfy the instability condition, 
$(E_1/\omega_1)(E_2/\omega_2)>0$.
We suppose that the mode 1 mentioned here is the upper frequency QPO of the twin HFQPOs.
It is interesting to note that the frequency ratio of the above two mode 1 oscillations
(the oscillation with $m_1=2$ and that of $m_1=3$) is close to 2 : 3 and depends little on
disk models (disk temperature) in the case of $a_*=0$, if they are reflected back around $r_{\rm in}$
(Lai and Tsang 2009, see also figure 7.2 of Kato (2016)). 

  
We should further notice that
there is the idea that the cause of HFQPOs in black-hole 
sources may be due to corotration amplification of the non-axisymmetric 
p-mode oscillations (\ref{resonance 5}) and (\ref{resonance 7}) 
(Tsang and Lai 2008, Lai and Tsang 2009) (notice that in figure 6 the corotation radius 
where $\omega=2\Omega$ is present just outside of the propagation region of $\omega_1$ oscillation). 

\begin{ack}   
We thank colleagues of M. Machida in the program of
high-resolution numerical simulations of black-hole accretion disks,
T. Kawashima, Y. Kudoh,  Y. Matsumoto, and R. Matsumoto,  
for their allowing us to use their unpublished results before publication. 
This research used computational resources of the K computer provided by the RIKEN Advances Institute for Computational Science 
through the HPCI System Research project 
(MM:Project ID:hp 180137, hp 190125).
This work was supported by JSPS KAKENHI Grant Number 19K03916.

\end{ack}

\bigskip\noindent
{\bf Appendix: Other Instabilities Described by the Wave-Wave 
Resonant Instability}

The wave-wave resonant instabilities resulting from 
other types of disk deformations are briefly summarized.
For simplicity, the cases of no magnetic fields are shown.

\bigskip\noindent
i) Warp resonant instability:

This is the case where the unperturbed disk is warped, i.e.,
\begin{equation}
        m_{\rm D}=-1, \  \quad n_{\rm D}=1, \quad {\rm and} \quad 
        \omega_{\rm D}\sim 0.
\label{deformation}
\end{equation}
For convenience, we have adopted here $m_{\rm D}=-1$  
without any physical changes by changing the sign of $\omega_{\rm D}$ simultaneously.
The set of resonant oscillations in this case is an axisymmetric g-mode oscillation and an one-armed  
p-mode oscillation:
\begin{eqnarray}
     &&{\rm mode}\ 1: \quad  {\rm g}-{\rm mode} \ {\rm oscillations\ with}\ 
                           (m_1=0 \ ,\ n_1=1, \ \omega_1<0), 
\label{append 2}               \\
     &&{\rm mode}\ 2: \quad  {\rm p}-{\rm mode} \ {\rm oscillations\ with}\
                           (m_2=1 \ ,\ n_2=0, \  \omega_2\sim -\omega_1>0),
\label{append 3}
\end{eqnarray}
where, for convenience, $\omega_1$ has been taken to be negative.
The propagation region of the former is specified by
$\omega_1^2-\kappa^2<0$, i.e., $-\kappa<\omega_1<\kappa$.
We take a trapped oscillation with $\omega_1<0$.
The propagation region of the latter oscillations
is specified by $(\omega_2-\Omega)^2>\kappa^2$, whose 
propagation region is divided into two parts.
What we are interested in is the oscillations in the region of
$\omega_2<\Omega-\kappa$ with $\omega_2>0$.

In mode 1 we have $E_1/\omega_1<0$, since oscillations are axisymmetric
and $\omega_1<0$.
In mode 2 we also have $E_2/\omega_2<0$, since the waves are inside the corotation resonance of $\omega_2=\Omega$ and $\omega_2>0$. 
The excitation condition 
(\ref{condition1}) is thus satisfied and the above coupled oscillations are
excited (Kato 2004, 2008; Ferreira and Ogilvie 2008).

\bigskip\noindent
ii) Eccentric resonant instability:

This is the case where the unperturbed disk is eccentric, i.e.,
\begin{equation} 
     m_{\rm D}=-1, \ \quad n_{\rm D}=0, \ \quad {\rm and} \ \quad 
        \omega_{\rm D}\sim 0.
\end{equation}
The set of oscillations which have resonant interaction are 
an axisymmetric g-mode oscillation 
and an one-armed g-mode oscillation:
\begin{eqnarray}
     &&{\rm mode}\ 1: \quad  {\rm g}-{\rm mode} \ {\rm oscillations\ with}\ 
                           (m_1=0 \ ,\ n_1=1, \ \omega_1<0), 
\label{append 2}               \\
     &&{\rm mode}\ 2: \quad  {\rm g}-{\rm mode} \ {\rm oscillations\ with}\
                           (m_2=1 \ ,\ n_2=1, \  \omega_2\sim -\omega_1>0).
\label{append 3}
\end{eqnarray}
The propagation region of the former oscillations is 
$-\kappa<\omega_1<\kappa$.
The propagation region of the latter oscillations
is $\Omega-\kappa<\omega_2<\Omega+\kappa$.

In the case of $\omega_1<0$, mode 1 has $E_1/\omega_1<0$.
On the other hand,
mode 2 with $\omega_2\sim -\omega_1>0$ has $E_2/\omega_2<0$,
if it is present in the propagation region inside the corotation radius of
$\omega_2=\Omega$.
Hence, the condition of resonant instability (\ref{condition1}) is satisfied. 
This resonant instability has been studied by Dewberry et. al. (2019a, b).
It should be noticed that the mode 2 considered above has the corotation 
radius within its propagation region, where the wave is damped
(Kato 2003, Li et al. 2003).
This might bring about no serious effect on the present resonant excitation.

\bigskip\noindent
iii) Tidal instability:

In tidally deformed disks, we have the deformation characterized by 
\begin{equation}
     m_{\rm D}=-3, \quad  n_{\rm D}=0, \quad {\rm  and} \quad \omega_{\rm D}=-3\Omega^*_{\rm obs},
\end{equation}
where $\Omega^*_{\rm obs}$ is the orbital frequency of the secondary star 
around the primary star.
We have again adopted here $m_{\rm D}=-3$ (and thus $\omega_{\rm D}<0$)
for convenience. 
A set of oscillations which have resonant interaction through this 
deformation are an one-armed p-mode oscillation 
($m_1=1$, $n_1=0$) and a two-armed p-mode oscillation ($m_2=2$,
$n_2=0$).
That is, 
\begin{eqnarray}
     &&{\rm mode}\ 1: \quad  {\rm p}-{\rm mode} \ {\rm oscillations\ with}\ 
                           (m_1=1 \ ,\ n_1=0, \ \omega_1>0), 
\label{append 2}               \\
     &&{\rm mode}\ 2: \quad  {\rm p}-{\rm mode} \ {\rm oscillations\ with}\
                           (m_2=2 \ ,\ n_2=0, \  \omega_2>0).
\label{append 3}
\end{eqnarray}
The propagation region of the former oscillations is characterized by
$(\omega_1-\Omega)^2>\kappa^2$.
This propagation region is separated into two:
$\omega_1>\Omega+\kappa$ and $\omega_1<\Omega-\kappa$.
We are interested in the oscillations in the latter propagation region
of $0<\omega_1<\Omega-\kappa$.
The propagation region of the two-armed p-mode oscillations are
characterized by $\omega_2>2\Omega+\kappa$ and $\omega_2<2\Omega-\kappa$.
We are interested in the oscillations in the latter propagation region.
This situation is shown in, for example, figure 12.1 by Kato (2016).

The former oscillation mode, mode 1, has $E_1/\omega_1<0$, since 
$\omega_1-\Omega<-\kappa$ and thus $\omega_1-\Omega<0$.
In the latter 
mode, mode 2, we have has $E_2/\omega_2<0$.
Hence, the instability condition (\ref{condition1}) is satisfied. 
This instability has been known as the
3:1 resonance in tidally deformed disks by simulations (Whitehurst 1988)
and by theoretical considerations 
(Hirose and Osaki 1990; Lubow 1991).

It is noted that many other types of tidal instabilities are possible,
since tidal waves are not limited to that considered above, if we extend our attention to cases where the orbit of the secondary star is eccentric and not 
coplanar (e.g., see Kato 2016). 

\end{document}